\documentclass{article}

\usepackage{amssymb}
\usepackage{amsmath}

\renewcommand{\title}[1]{
\begin{center} \Large \bf #1 \end{center}
}

\renewcommand{\author}[3]{
 \begin{center} #1 \\ \ \\
  {\it #2} \\ \ \\
  {\texttt #3}
 \end{center}
}

\begin{document}

\begin{titlepage}

\begin{flushleft}
February 2004
\end{flushleft}

\begin{flushright}
OCHA-PP-222
\end{flushright}

\vspace{10mm}

\title{PST-type $SL(2;\mbox{\bf R})$-covariant Super D3-brane Action \\
in Flat Spacetime}

\vspace{5mm}

\author
{Toshiya Suzuki}
{Department of Physics, Faculty of Science, Ochanomizu University\\
Institute of Humanities and Sciences, Ochanomizu University\\
2-1-1 Otsuka, Bunkyo-ku, Tokyo 112-8610, Japan}
{tsuzuki@phys.ocha.ac.jp}

\vspace{5mm}

\abstract{
We give an explicit form of the PST-type $SL(2;\mbox{\bf R})$-covariant
 super D3-brane action for the flat Minkowski background. To this end, 
we follow the prescription developed by Hatsuda and Kamimura. 
As an application of the action, we obtain the supercharge of the action
 by using the standard Noether's method and calculate the Poisson
 bracket algebra of the supercharge. The central charge of the
 supersymmetry algebra is given in a manifestly $SL(2;\mbox{\bf R})$-covariant way.
}

\end{titlepage}

%

The D3-brane is a non-perturbative excitation of the type-IIB superstring
theory. 
It has two important properties; (i) it is a BPS saturated object, (ii)
it is invariant under the $SL(2;\mbox{\bf R})$ self-duality \cite{T,GG,APPS}
\footnote{On quantum level, the $SL(2;\mbox{\bf R})$ 
self-duality becomes discrete $SL(2;\mbox{\bf Z})$. However, all
analyses are performed classically, so we use the term $SL(2;\mbox{\bf
R})$ throughout this letter.}, which is an important symmetry of the
type-IIB superstring theory \cite{HT}. 

There are several attempts to construct the type-IIB superstring theory
in a manifestly $SL(2;\mbox{\bf R})$-covariant way or to clarify the
origin of the $SL(2;\mbox{\bf R})$ self-duality. One of them is known as
the F-theory, which is an idea that the duality arises from $12$
dimensional geometry \cite{V}.
At present, we do not know final answer to the question that these
attempts succeed or not, however, if do, the D3-brane will play an
important role in these attempts and it will be useful to formulate the
D3-brane in a manifestly $SL(2;\mbox{\bf R})$-covariant way.

In order to formulate the D3-brane manifestly $SL(2;\mbox{\bf
R})$-covariantly, one has to introduce  $SL(2;\mbox{\bf R}$ doublet
vector fields, which propagate on the D3-brane world volume.
Consequently, one has redundant degrees of freedom.
Using the formalism developed by Pasti, Sorokin and Tonin
\cite{PST1,PST2,PST3}, Nurmagambetov resolved this problem and gave a
manifestly $SL(2;\mbox{\bf R})$-covariant action of the bosonic D3-brane
\cite{N} (See also \cite{PST4,BLNPST,B1,B2,BNS}). After this, in
\cite{TS}, the author extended his work to supersymmetric
version and proved the $\kappa$-symmetry of the supersymmetric action
, which is known as a main consistency condition for actions of
supersymmetric extended objects. 

However, the work in \cite{TS} has a drawback; there, the action is not given
in explicit form. 
In order to proof the $\kappa$-symmetry, we need
so-called supergravity constraints, equivalent to the field equations
of targetspace supergravity. They are algebraic constraints on
components of the supergravity field strength, for example, 
$R^{(5)}$ in the type IIB case.
On the other hand, the Wess-Zumino term in the action which, roughly speaking,
represents coupling to the background fields takes the following form,
\begin{equation} 
L_{WZ} \sim C^{(4)} , \nonumber 
\end{equation}
where $C^{(4)}$ is the background $4$ form field and $d C^{(4)} \sim
R^{(5)}$. So, in fact, we only know the form of $d L_{WZ}$.
 
In order to give the explicit $L_{WZ}$, we must solve the field equation
of supergravity. Of course, it is difficult task for general cases, but
for some ones it is possible. The simplest and very important is
the case of the Minkowski flat background. In the superstring theory and
also in the supermembrane one, the analysis of the action in the flat
background is the most basic and yields many important results.       

In this article, as preparation to future research, we will give 
the explicit form of the super D3-brane action in the flat background.
To obtain the explicit form, we follow the method developed in
\cite{HK1,HK2} with a bit device to treat the action in a way that keeps
manifest $SL(2;\mbox{\bf R})$-covariance.
We will also investigate the Poisson bracket algebra of the
supersymmetry as an application and then give the central charge of the
supersymmetry algebra in manifestly $SL(2;\mbox{\bf R})$-covariant form.
         
%

%
To begin with, we review the PST-type $SL(2;\mbox{\bf R})$-covariant
D3-brane action \cite{N,TS}. It is composed of 3 parts
\begin{equation}
S = S_{DBI} +S_{PST} + S_{WZ} .
\label{s123}
\end{equation}
The first part is called Dirac-Born-Infeld (DBI) term and given as
\begin{eqnarray}
S_{DBI} &=& \int d^{4} \xi - \sqrt{-g} f , \nonumber \\
f &=& \sqrt{1 + {\tilde H}_{i , r} (\epsilon M \epsilon)^{r s} {{
\tilde H}^{i}}_{, s} + \frac{1}{2} ({\tilde H}_{i , r} \epsilon^{r s} {\tilde H}_{j , s}) ({{\tilde H}^{i}}_{, t} \epsilon^{t u} {{\tilde H}^{j}}_{, u})} .
\label{DBI}
\end{eqnarray}
The second part, which is called Pasti-Sorokin-Tonin (PST) term, is given by
\begin{equation}
S_{PST} = \int d^{4} \xi \sqrt{-g} \frac{1}{2} {\tilde H}_{i , r} \epsilon^{r s} {H^{i j}}_{, s} v_{j} ,
\label{PST}
\end{equation}
and the third, which is called Wess-Zumino (WZ) term, is given by
\begin{equation}
S_{WZ} = \int C^{(4)} - \frac{1}{2} H^{(2)}_{r} C^{(2)}_{s} \epsilon^{r s} .
\label{WZ} 
\end{equation}
They are described by the superembedding fields $z^M =
(x^m,\theta^{\alpha},{\bar \theta}^{\dot \alpha})$, the $SL(2;\mbox{\bf R})$
doublet vector fields $A_{i,r} , (r=1,2)$ and the auxiliary scalar field $a$.
$A_{i,r}$ are often called the BI fields and $a$ is the PST field.
Definitions of quantities appearing in (\ref{DBI}),(\ref{PST}),(\ref{WZ})
are as follows.

\noindent
$\bullet$ \ The induced metric $g_{i j}$ is defined by
\begin{equation}
g_{i j} = \pi^{m}_{i} \pi^{n}_{j} \eta_{m n} ,
\label{g}
\end{equation}
where $\pi^{m}_i$ is the worldvolume pullback of $\pi^{m}$
\begin{equation}
\pi^{m} =  d x^{m} + Re(d \bar{\theta} \Gamma^{m} \theta) ,
\label{pi}
\end{equation}
which is the bosonic part of the supervielbein 1-form $E^{A}=(\pi^{m}, d
\theta^{\alpha}, d {\bar \theta}^{\dot \alpha})$ for the flat superspace.

\noindent
$\bullet$ \ ${\tilde H}_{i , r}$ is defined by
\begin{equation}
{\tilde H}_{i , r} = v^{j} *H_{j i , r} \ , \ {*H^{i j}}_{, r} = \frac{\epsilon^{i j k l}}{2! \sqrt{-g}} H_{k l , r} .
\label{tilH}
\end{equation}
$H_{ij,r}$ is called the improved field strength and is
defined by 
\begin{equation}
H_{i j , r} = F_{ij , r} - C_{ij , r} ,
\label{H}
\end{equation}
where $F_{ij , r} = \partial_{i} A_{j , r} - \partial_{j} A_{i , r}$ 
and $C_{ij , r}$ is the worldvolume pullback of the spacetime
2-form $C^{(2)}_{,r}$ .
Also $v^i$ is the normalized gradient vector of $a$
\begin{equation}
v^{i} = \frac{\partial^{i} a}{\sqrt{- (\partial a)^{2}}} \ , \ (v)^{2} = -1 .
\label{v}
\end{equation}

\noindent
$\bullet$ \ $M$ is the $SL(2;\mbox{\bf R})$ matrix and is written in terms
of the spacetime scalars, dilaton $\phi$ and axion $\chi$
\begin{equation}
M = M_{r s} = \frac{1}{e^{-\phi}}
\left[
\begin{array}{cc}
1 & \chi \\
\chi & \chi^{2} + e^{-2 \phi}
\end{array}
\right] .
\label{M}
\end{equation}
We suppose $\phi$ and $\chi$ are constant. And
$\epsilon_{rs} = \epsilon^{rs} = \left[
\begin{array}{cc}
0 & 1 \\
-1 & 0
\end{array}
\right] $ is the $SL(2;\mbox{\bf R})$-invariant tensor.

\noindent
$\bullet$ \ $C^{(4)}$ is the background 4-form with self-dual 5-form
field strength and $C^{(2)}_{,r}$ is the $SL(2;\mbox{\bf R})$ doublet background 2-form.  
Even in flat Minkowski background, they can have non-zero components.

%
At this stage we briefly comment on the $\kappa$ symmetry and the PST
symmetries, which are local symmetries of the action.
In general, actions of supersymmetric extended objects must possess 
the $\kappa$ symmetry as their consistency condition.
The proof of the $\kappa$ symmetry for the PST-type $SL(2;\mbox{\bf
R})$-covariant super D3-brane action was done in \cite{TS} for general
on shell type IIB supergravity background. Flat Minkowski spacetime with
constant scalars is a special case of on shell
background, so in this letter we do not present the proof of $\kappa$
symmetry.

On the other hand, the PST symmetries are a set of two types of symmetries.
One is required in order to reduce the equation of
motion for $A_{i , r}$ to the non-linear electro-magnetic relation
between $SL(2;\mbox{\bf R})$ doublet BI fields $A_{i , r}$. 
The other is needed to gauge away the auxiliary field $a$. 

%
Before explaining the supersymmetry of the action, 
we rewrite the action into more convenient form for later
analysis.
We describe the background scalars $\phi$ and $\chi$ in terms of the complex
$SL(2;\mbox{\bf R})$ doublet scalars $u^r$ and ${\bar u}^r$ \cite{SW,S,HW},
\begin{equation}
\tau = - \frac{u^1}{u^2} \ , \ \tau = \chi + i e^{- \phi} , 
\end{equation}
where $u^r$ and ${\bar{u}^r}$ are constrained by
\begin{equation}
\frac{i}{2} \epsilon_{r s} u^r {\bar u}^s = 1 .
\label{euu}
\end{equation} 
We assign the global $U(1)$ charge $+1$ or $-1$ for $u^r$ or ${\bar
u}^r$.
Also we assign $+\frac{1}{2}$ or $-\frac{1}{2}$ for $\theta$ or ${\bar
\theta}$.
 
Using $u$'s we contract the indices of the $SL(2;\mbox{\bf R})$ doublet
quantities, for example 
\begin{equation}
A_i = u^r A_{i , r} \ , \ {\bar A}_i = {\bar u}^r A_{i , r} , 
\label{uA}
\end{equation}
\begin{equation}
F_{ij} = u^r F_{ij , r} \ , \ {\bar F}_{ij} = {\bar u}^r F_{ij , r} ,
\label{uF}
\end{equation}
and so on. For constant scalar backgrounds, 
\begin{equation}
F_{ij} = \partial_i A_j - \partial_j A_i \ , \ {\bar F}_{ij} = \partial_i {\bar A}_j - \partial_j {\bar A}_i .
\label{uFduA}
\end{equation}

%
In terms of those $SL(2;\mbox{\bf R})$ singlet but $U(1)$ charged
quantities,
the action is represented by
\begin{eqnarray}
S_{DBI} &=& \int d^{4} \xi - \sqrt{-g} f , \nonumber \\
f &=& \sqrt{1 + \bar{\tilde H}_{i} {\tilde H}^{i} + \frac{1}{2} Im(\bar{\tilde H}_{i} {\tilde H}_{j}) Im(\bar{\tilde H}^{i} {\tilde H}^{j})} ,
\label{DBI2}
\end{eqnarray}
\begin{equation}
S_{PST} = \int d^{4} \xi \sqrt{-g} \frac{1}{2} Im(\bar{\tilde H}_{i} H^{i j}) v_{j} ,
\label{PST2}
\end{equation}
and
\begin{equation}
S_{WZ} = \int C^{(4)} - \frac{1}{2}Im(\bar{H}^{(2)} C^{(2)}) .
\label{WZ2}
\end{equation}

%
To give explicit form of the action, we must know explicit form of
$C^{(2)}$ , ${\bar C}^{(2)}$ and $C^{(4)}$, but in general we know only
their field strength
\begin{equation}
R^{(3)} = d C^{(2)} \ , \ \bar{R}^{(3)} = d \bar{C}^{(2)} , 
\label{R3dC2}
\end{equation}
and
\begin{equation}
R^{(5)} = d C^{(4)} - \frac{1}{2} Im(\bar{C}^{(2)} R^{(3)}) .
\label{R5dC4}
\end{equation}
For flat Minkowski background they are \cite{TS,CW} \footnote{
In the standard non-manifestly $SL(2;\mbox{\bf R})$-covariant formalism,
the explicit forms of $C^{(4)}$'s are known \cite{CvGNW,HK1}. But for our
purpose, they are not convenient.
},
\begin{equation}
R^{(3)} = \pi \cdot d \theta \Gamma d \theta \ , \ \bar{R}^{(3)} = \pi \cdot d \bar{\theta} \Gamma d \bar{\theta} ,
\label{R3}
\end{equation}
and
\begin{equation}
R^{(5)} = \frac{1}{2i} (d \bar{\theta} \frac{\pi^{3} \!\!\!\!\!\! /}{3!} d \theta - d \theta \frac{\pi^{3} \!\!\!\!\!\! /}{3!} d \bar{\theta}) .  
\label{R5}
\end{equation}

%
Let us explain the supersymmetry of the action.
The supersymmetry transformation is defined by the background isometry,
that is, such a translation that does not change the background
supervielbein 1-form $E^A$ and the background field strength $R^{(3)}$ ,
${\bar R}^{(3)}$ and $R^{(5)}$. 
For flat Minkowski case (\ref{R3}),(\ref{R5}), the supersymmetry
transformations are given by
\begin{equation}
\delta \theta = \varepsilon \ , \ \delta \bar{\theta} = \bar{\varepsilon} ,
\label{stheta}
\end{equation}
and
\begin{equation}
\delta x^{m} = - Re(\bar{\theta} \Gamma^{m} \varepsilon) .
\label{sx}
\end{equation}
One can check the invariance of 
$E^A = (\pi^m , d \theta^{\alpha} , d \theta^{\bar \alpha})$ under
(\ref{stheta}),(\ref{sx}). 
For $A_{i}$ or ${\bar A}_i$, the supersymmetry transformation is defined
so that the improved field strength $H_{ij}$ or ${\bar H}_{ij}$
keeps to be invariant.
For (\ref{R3}), we have  
\begin{eqnarray}
C^{(2)} &=& (\pi - d \theta \Gamma \bar{\theta}) \cdot d \theta \Gamma \theta , \nonumber \\
\bar{C^{(2)}} &=& (\pi - d \bar{\theta} \Gamma \theta) \cdot d \bar{\theta} \Gamma \bar{\theta} ,
\label{C2}
\end{eqnarray}
up to total derivatives, then obtain
\begin{eqnarray}
\delta A^{(1)} &=& \ \ d x \cdot \theta \Gamma \varepsilon \nonumber \\
 & & + \frac{1}{6} (\varepsilon \Gamma \bar{\theta} \cdot d \theta \Gamma \theta + \varepsilon \Gamma \theta \cdot d \theta \Gamma \bar{\theta} - 2 \varepsilon \Gamma \theta \cdot d \bar{\theta} \Gamma \theta -2 \bar{\varepsilon} \Gamma \theta \cdot d \theta \Gamma \theta) , \nonumber \\
\delta \bar{A}^{(1)} &=& \ \ d x \cdot \bar{\theta} \Gamma \bar{\varepsilon} \nonumber \\
 & & + \frac{1}{6} (\bar{\varepsilon} \Gamma \theta \cdot d \bar{\theta} \Gamma \bar{\theta} + \bar{\varepsilon} \Gamma \bar{\theta} \cdot d \bar{\theta} \Gamma \theta - 2 \bar{\varepsilon} \Gamma \bar{\theta} \cdot d \theta \Gamma \bar{\theta} -2 \varepsilon \Gamma \bar{\theta} \cdot d \bar{\theta} \Gamma \bar{\theta}) .
\label{sA}
\end{eqnarray}
For $a$, we set 
\begin{equation}
\delta a = 0 .
\label{sa}
\end{equation}
Under (\ref{stheta}),(\ref{sx}),(\ref{sA}),(\ref{sa}), the action is
invariant up to total derivatives. 

%

Now we integrate $d L_{WZ} = R^{(5)} - \frac{1}{2}Im(\bar{H}^{(2)}
R^{(3)})$ to obtain the explicit form of the action.
To put it through, we follow the prescription developed in
\cite{HK1,HK2} (See also \cite{APS1,APS2}).
We rewrite $d L_{WZ}$ into the following form,
\begin{equation}
d L_{WZ} = - \frac{1}{2i} d \bar{\Theta} K e^{-I \frac{H^{(2)} + \bar{H}^{(2)}}{2}} e^{- \pi \!\!\! /} d \Theta ,
\label{dWZ}
\end{equation}
on the understanding that in R.H.S. only 5-form and $U(1)$
neutral terms are taken \footnote{In the rest, the analogous rules 
are understood wherever formal sums of forms with different degrees and 
those of quantities with different $U(1)$ charges appear.}. Recall $H$
or ${\bar H}$ has $U(1)$ charge $+1$ or $-1$ and $\theta$ or ${\bar
\theta}$ has $+\frac{1}{2}$ or $-\frac{1}{2}$. 
Using the formula (see Appendix \ref{s.pe}),
\begin{equation}
d \ [ K e^{- I \frac{H + \bar{H}}{2}} e^{\pm {\pi \!\!\! /}} ] = (\pm \frac{1}{4} d {V \!\!\!\! /}) K e^{- I \frac{H + \bar{H}}{2}} e^{\mp {\pi \!\!\! /}} + K e^{- I \frac{H + \bar{H}}{2}} e^{\pm {\pi \!\!\! /}} (\pm \frac{1}{4} d {V \!\!\!\! /}) ,
\label{dKeIHepi}
\end{equation}
and Eq.(\ref{dVdTheta}) in Appendix \ref{ss.fi} 
and Eq.(\ref{dOmega}) in Appendix \ref{ss.d}, 
we can integrate $d L_{WZ}$ and then obtain 
\begin{equation}
L_{WZ} = - \frac{1}{2i} d \bar{\Theta} K e^{-I \frac{H^{(2)} + \bar{H}^{(2)}}{2}} e^{- \pi \!\!\! /} \Omega^{(+)} .
\label{WZE}
\end{equation}

%
Here, we would like to stress the virtue of the global $U(1)$
symmetry. To keep the manifest $SL(2;\mbox{\bf R})$-covariance requires to
treat unbarred quantities (for example $H$) and barred ones (for
example ${\bar H}$) symmetrically as possible. 
So we use the method of description like Eqs.(\ref{dKeIHepi}),(\ref{WZE}), 
however, they have many redundant terms.
In order to extract proper terms from them, the global
$U(1)$ symmetry plays essential role. 
  
%

As an application of the action with the explicit WZ term (\ref{WZE}), 
we investigate the supersymmetry algebra \cite{ST,H,HK1,HK2}.
%
Following the standard Noether's method, the supersymmetry charges are
given as
\begin{equation}
 \varepsilon Q + \bar{\varepsilon} \bar{Q} =  \varepsilon \{ Q^1 + Q^2 \} + \bar{\varepsilon} \{ \bar{Q}^1 + \bar{Q}^2 \} ,
\label{Q}
\end{equation}
where
\begin{eqnarray}
\varepsilon Q^1 + \bar{\varepsilon} \bar{Q}^1 &=& \int d^3 \xi \left[ \delta x^m p_m + \varepsilon \zeta + \bar{\varepsilon} \bar{\zeta} + \delta A_a E^a + \delta \bar{A}_a \bar{E}^a \right] \nonumber \\
 &=& \int d^3 \xi 
\left[
\begin{array}{l}
\ \ \varepsilon \left\{
\begin{array}{l}
\frac{1}{2} \Gamma \bar{\theta} \cdot p + \zeta - \Gamma \theta \cdot \partial_a x E^a \\
+ \frac{1}{6} \Gamma \bar{\theta} \cdot \partial_a \theta \Gamma \theta E^a + \frac{1}{6} \Gamma \theta \cdot \partial_a \theta \Gamma \bar{\theta} E^a \\
- \frac{1}{3} \Gamma \theta \cdot \partial_a \bar{\theta} \Gamma \theta E^a - \frac{1}{3} \Gamma \bar{\theta} \cdot \partial_a \bar{\theta} \Gamma \bar{\theta} \bar{E}^a 
\end{array}
\right\} \\
+ \bar{\varepsilon} \left\{
\begin{array}{l}
\frac{1}{2} \Gamma \theta \cdot p + \bar{\zeta} - \Gamma \bar{\theta} \cdot \partial_a x \bar{E}^a \\
+ \frac{1}{6} \Gamma \theta \cdot \partial_a \bar{\theta} \Gamma \bar{\theta} \bar{E}^a + \frac{1}{6} \Gamma \bar{\theta} \cdot \partial_a \bar{\theta} \Gamma \theta \bar{E}^a \\
- \frac{1}{3} \Gamma \bar{\theta} \cdot \partial_a \theta \Gamma \bar{\theta} \bar{E}^a - \frac{1}{3} \Gamma \theta \cdot \partial_a \theta \Gamma \theta E^a 
\end{array}
\right\}
\end{array}
\right] ,
\label{Q1}
\end{eqnarray}
where $p_m$,$\zeta$,${\bar \zeta}$,$E^a$,${\bar E}^a$ are canonical
momenta of $x^m$,$\theta$,${\bar \theta}$,$A_a$,${\bar A}_a$ respectively.
Also
\begin{equation}
\varepsilon Q^2 + \bar{\varepsilon} \bar{Q}^2 = - \int d^3 \xi [ q ] ,
\label{Q2}
\end{equation}
where $[ q ]$ means $d \xi^1 d \xi^2 d \xi^3$ coefficient of 3-form
$q$, which is related to the supersymmetry variation of the WZ
term under the following relation
\begin{equation}
\delta L_{WZ} = d q .
\label{dWZdq}
\end{equation}
Using (\ref{WZE}) and Eq. (\ref{deltaOmega}) in Appendix \ref{ss.delta},
we obtain 
\begin{eqnarray}
q &=& - \frac{1}{2i} \bar{\Omega}^{(-)} K e^{-J \frac{H + \bar{H}}{2} } e^{+ \pi \!\!\! / } \Xi \nonumber \\
 & & - \frac{1}{2i} d \bar{\Theta} K e^{-J \frac{H + \bar{H}}{2} } e^{- \pi \!\!\! / } \Sigma^{(-)}_{\Xi} \nonumber \\
 & & + \frac{1}{4i} \bar{\Omega}^{(-)} K e^{-J \frac{H + \bar{H}}{2} } e^{+ \pi \!\!\! / } {V  \!\!\!\! /}_{ d \Theta , \Xi } \Omega^{(-)} ,
\label{q}
\end{eqnarray}
where
\begin{equation}
\Xi = \left[
\begin{array}{c}
\varepsilon \\
{\bar \varepsilon}
\end{array}
\right] \ , \ {\bar \Xi} = [ {\bar \varepsilon} , \varepsilon ] .
\label{Xi}
\end{equation}

%
Then we can calculate Poisson bracket algebra of supercharge
(\ref{Q}), 
\begin{eqnarray}
& &
\left[
\begin{array}{cc}
\{ \bar{\varepsilon} \bar{Q} \ , \ \varepsilon' Q \}_{PB} &  
\{ \bar{\varepsilon} \bar{Q} \ , \ \bar{\varepsilon}' \bar{Q} \}_{PB} \\
\{ \varepsilon Q \ , \ \varepsilon' Q \}_{PB} &
\{ \varepsilon Q \ , \ \bar{\varepsilon}' \bar{Q} \}_{PB}
\end{array}
\right]
\nonumber \\
&=&
\left[
\begin{array}{cc}
\{ \bar{\varepsilon} \bar{Q}^1 \ , \ \varepsilon' Q^1 \}_{PB} & 
\{ \bar{\varepsilon} \bar{Q}^1 \ , \ \bar{\varepsilon}' \bar{Q}^1 \}_{PB} \\
\{ \varepsilon Q^1 \ , \ \varepsilon' Q^1 \}_{PB} & 
\{ \varepsilon Q^1 \ , \ \bar{\varepsilon}' \bar{Q}^1 \}_{PB}
\end{array}
 \right]
\nonumber \\
&+&
\left[
\begin{array}{cc}
\{ \bar{\varepsilon} \bar{Q}^1 \ , \ \varepsilon' Q^2 \}_{PB} & 
\{ \bar{\varepsilon} \bar{Q}^1 \ , \ \bar{\varepsilon}' \bar{Q}^2 \}_{PB} \\
\{ \varepsilon Q^1 \ , \ \varepsilon' Q^2 \}_{PB} & 
\{ \varepsilon Q^1 \ , \ \bar{\varepsilon}' \bar{Q}^2 \}_{PB}
\end{array}
 \right]
+
\left[
\begin{array}{cc}
\{ \bar{\varepsilon} \bar{Q}^2 \ , \ \varepsilon' Q^1 \}_{PB} & 
\{ \bar{\varepsilon} \bar{Q}^2 \ , \ \bar{\varepsilon}' \bar{Q}^1 \}_{PB} \\
\{ \varepsilon Q^2 \ , \ \varepsilon' Q^1 \}_{PB} & 
\{ \varepsilon Q^2 \ , \ \bar{\varepsilon}' \bar{Q}^1 \}_{PB}
\end{array}
\right]
\label{QQ}
\end{eqnarray}
according to the classical canonical relations
\begin{eqnarray}
\{ p_m(\xi) \ , \ x^n(\xi') \}_{PB} &=& \delta_m^n\delta^3(\xi - \xi') , \nonumber \\
\{ \zeta_{\alpha}(\xi) \ , \ \theta^{\beta}(\xi') \}_{PB} &=& \delta_{\alpha}^{\beta} \delta^3(\xi - \xi') , \nonumber \\
\{ {\bar \zeta}_{\dot \alpha}(\xi) \ , \ {\bar \theta}^{\dot \beta}(\xi') \}_{PB} &=& \delta_{\dot \alpha}^{\dot \beta} \delta^3(\xi - \xi') , \nonumber \\
\{ E_a(\xi) \ , \ A_b(\xi') \}_{PB} &=& \delta_a^b \delta^3(\xi - \xi') , \nonumber \\
\{ {\bar E}_a(\xi) \ , \ {\bar A}^b(\xi') \}_{PB} &=& \delta_a^b \delta^3(\xi - \xi') .
\label{ccr}
\end{eqnarray}
The results are
\begin{eqnarray}
& &
\left[
\begin{array}{cc}
\{ \bar{\varepsilon} \bar{Q}^1 \ , \ \varepsilon' Q^1 \}_{PB} & 
\{ \bar{\varepsilon} \bar{Q}^1 \ , \ \bar{\varepsilon}' \bar{Q}^1 \}_{PB} \\
\{ \varepsilon Q^1 \ , \ \varepsilon' Q^1 \}_{PB} & 
\{ \varepsilon Q^1 \ , \ \bar{\varepsilon}' \bar{Q}^1 \}_{PB}
\end{array}
\right] \nonumber \\
&=&
\int d^3 \sigma
\left[
\begin{array}{cc}
\bar{\varepsilon} \left(
\begin{array}{l}
- \Gamma \cdot p \\
+ \frac{1}{48} \Gamma_{l m n} \cdot \theta \Gamma^{l m n} \theta \ \partial_a E^a \\
- \frac{1}{48} \Gamma_{l m n} \cdot \bar{\theta} \Gamma^{l m n} \bar{\theta} \ \partial_a \bar{E}^a 
\end{array}
\right) \varepsilon' &
\bar{\varepsilon} \left(
\begin{array}{l}
+ 2 \Gamma \cdot \partial_a x \ \bar{E}^a \\
- \frac{2}{3} \Gamma \cdot \theta \Gamma \partial_a \bar{\theta} \ \bar{E}^a \\
- \frac{1}{2} \Gamma \cdot \theta \Gamma \bar{\theta} \ \partial_a \bar{E}^a
\end{array}
\right) \bar{\varepsilon}' \\
\varepsilon \left(
\begin{array}{l}
+ 2 \Gamma \cdot \partial_a x \ E^a \\
- \frac{2}{3} \Gamma \cdot \bar{\theta} \Gamma \partial_a \theta \ E^a \\
- \frac{1}{2} \Gamma \cdot \bar{\theta} \Gamma \theta \ \partial_a E^a
\end{array}
\right) \varepsilon' &
\varepsilon \left(
\begin{array}{l}
- \Gamma \cdot p \\
+ \frac{1}{48} \Gamma_{l m n} \cdot \bar{\theta} \Gamma^{l m n} \bar{\theta} \ \partial_a \bar{E}^a \\
- \frac{1}{48} \Gamma_{l m n} \cdot \theta \Gamma^{l m n} \theta \ \partial_a E^a
\end{array}
\right) \bar{\varepsilon}'
\end{array}
\right]
\label{Q1Q1}
\end{eqnarray}
and
\begin{eqnarray}
& & \left[
\begin{array}{cc}
\{ \bar{\varepsilon} \bar{Q}^1 \ , \ \varepsilon' Q^2 \}_{PB} & 
\{ \bar{\varepsilon} \bar{Q}^1 \ , \ \bar{\varepsilon}' \bar{Q}^2 \}_{PB} \\
\{ \varepsilon Q^1 \ , \ \varepsilon' Q^2 \}_{PB} & 
\{ \varepsilon Q^1 \ , \ \bar{\varepsilon}' \bar{Q}^2 \}_{PB}
\end{array}
 \right]
+
( 1 \leftrightarrow 2 ) \nonumber \\
&=& - \int d [ \delta_{ 1 } q_{ 2 }  - ( 1 \leftrightarrow 2) ] \nonumber \\
&=& \int \frac{1}{4i} (2 \bar{\varXi} - \bar{\Omega}^{(-)} {V \!\!\!\! /}_{d \Theta , \varXi}) K e^{ -I \frac{H + \bar{H}}{2} } e^{+ \pi \!\!\! /} (2 \varXi' - {V \!\!\!\! /}_{d \Theta , \varXi'} \Omega^{(-)}) ,
\label{Q1Q2}
\end{eqnarray}
up to total derivatives. To obtain these results, we use
Eqs.(\ref{deltaOmega}),(\ref{deltaSigma}) in Appendix \ref{ss.delta}

For bosonic configuration, that is $\theta = {\bar \theta} = 0$,
the supersymmetry algebra reduces to 
\begin{eqnarray}
& &
\left[
\begin{array}{cc}
\{ \bar{\varepsilon} \bar{Q} \ , \ \varepsilon' Q \}_{PB} & 
\{ \bar{\varepsilon} \bar{Q} \ , \ \bar{\varepsilon}' \bar{Q} \}_{PB} \\
\{ \varepsilon Q \ , \ \varepsilon' Q \}_{PB} & 
\{ \varepsilon Q \ , \ \bar{\varepsilon}' \bar{Q} \}_{PB}
\end{array}
\right] \nonumber \\
&=&
\int d^3 \sigma
\left[
\begin{array}{cc}
\bar{\varepsilon} \left(
\begin{array}{l}
- \Gamma \cdot p \\
- \frac{1}{3! i} \Gamma_{lmn} \varepsilon^{abc} \partial_a x^l \partial_b x^m \partial_c x^n 
\end{array}
\right) \varepsilon' &
\bar{\varepsilon} \left(
\begin{array}{l}
+ 2 \Gamma \cdot \partial_a x \ \bar{E}^a \\
+ \frac{1}{2 i} \Gamma \cdot \partial_a x \ \varepsilon^{abc} \partial_b A_c
\end{array}
\right) \bar{\varepsilon}' \\
\varepsilon \left(
\begin{array}{l}
+ 2 \Gamma \cdot \partial_a x \ E^a \\
- \frac{1}{2 i} \Gamma \cdot \partial_a x \ \varepsilon^{abc} \partial_b \bar{A}_c
\end{array}
\right) \varepsilon' &
\varepsilon \left(
\begin{array}{l}
- \Gamma \cdot p \\
+ \frac{1}{3! i} \Gamma_{lmn} \varepsilon^{abc} \partial_a x^l \partial_b x^m \partial_c x^n
\end{array}
\right) \bar{\varepsilon}'
\end{array}
\right] .
\label{QQb}
\end{eqnarray}
Eqs.(\ref{Q1Q1}),(\ref{Q1Q2}) and Eq.(\ref{QQb}) are manifestly
$SL(2;\mbox{\bf R})$-covariantization of the results obtained in
\cite{HK1}.

%

In this letter, we gave the explicit form of the PST-type
$SL(2;\mbox{\bf R})$-covariant super D3-brane action in the flat
Minkowski background by using the prescription of \cite{HK1,HK2}.
By virtue of the global $U(1)$ symmetry, the calculation was simplified. 
Then, following the standard Noether's method we obtained the
supercharge and calculated the Poisson bracket algebra of the
supercharge. The results are the $SL(2;\mbox{\bf R})$-covariantization of
those of \cite{HK1,HK2}.

In \cite{HK1,HK2}, the canonical constraint analysis was also performed,
and showed one half of the constraints associated with the
$\kappa$-symmetry belong to the first class and the other half to the
second class, and the other local symmetries' constraints are all the
first class ones.
In order to do the similar analysis for our action, however, we need to
know about the constraints associated with the PST symmetries.
As seen from Eqs. (\ref{DBI}),(\ref{PST}), the dependence of the PST
field $a$ is complicated.
So to obtain the constraint relation associated with the PST symmetries 
needs some efforts. 
We hope to report the results of this analysis in the forthcoming paper
\cite{TSf}.  

%

\ \ \ \ \

\ \ \ \ \

\noindent
{\Large \bf Appendix}

\appendix

\section{Spinors and $\Gamma$ matrices }
\label{s.gm}

$\theta^{\alpha}$ and ${\bar \theta}^{\dot \alpha}$
denote $10$dim. Weyl spinors, then they have net complex 16 components.
However we take a real representation,
so $\alpha$ and ${\dot \alpha}$ run from $1$ to $32$ and $\Gamma$
matrices are real $32 \times 32$ matrices.
In this case, undotted indices $\alpha$ and dotted indices ${\dot \alpha}$ are
equivalent.

\subsection{index position}
\label{ss.id}

The standard index position of $\Gamma$ matrices is set to be 
${\Gamma^{\alpha}}_{\beta}$, but one often encounters
$\Gamma$ matrices like $\Gamma_{\alpha \beta}$.
They mean that
\begin{equation} 
\Gamma_{\alpha \beta} = C_{\alpha \gamma} {\Gamma^{\gamma}}_{\beta} \ , \ C = \Gamma_0 .
\label{is}
\end{equation}
Their symmetry properties are, for example, 
\begin{eqnarray}
{\Gamma^k}_{\alpha \beta} &:& symmetric , \nonumber \\
{\Gamma^{kl}}_{\alpha \beta} &:& symmetric , \nonumber \\
{\Gamma^{klm}}_{\alpha \beta} &:& anti-symmetric , \nonumber \\
{\Gamma^{klmn}}_{\alpha \beta} &:& anti-symmetric .
\label{Gammasym} 
\end{eqnarray}
where
\begin{eqnarray}
\Gamma^{kl} &=& \frac{1}{2} \Gamma^{[ k} \Gamma^{l ]} , \nonumber \\
\Gamma^{klm} &=& \frac{1}{3!} \Gamma^{[ k} \Gamma^{l} \Gamma^{m ]} , \nonumber \\
\Gamma^{klmn} &=& \frac{1}{4!} \Gamma^{[ k} \Gamma^{l} \Gamma^{m} \Gamma^{n ]} .
\label{Gammas}
\end{eqnarray}

\subsection{{\it double-decker} notation}
\label{ss.ddn}

We often use {\it double-decker} notation 
\begin{equation}
\Theta = \left[
\begin{array}{c}
\theta \\
\bar{\theta}
\end{array}
\right] \ , \ \bar{\Theta} = \left[
\begin{array}{cc}
\bar{\theta} & \theta
\end{array}
\right] ,
\label{double}
\end{equation}
and it is useful to define the following $2 \times 2$ matrices
\begin{equation}
{\it 1} = \left[
\begin{array}{cc}
1 & 0 \\
0 & 1
\end{array}
\right] \ , 
\ I = \left[
\begin{array}{cc}
0 & 1 \\
1 & 0
\end{array}
\right] \ , 
\ J = \left[
\begin{array}{cc}
0 & -i \\
i & 0
\end{array}
\right] \ ,
\ K = \left[
\begin{array}{cc}
1 & 0 \\
0 & -1
\end{array}
\right] .
\label{1IJK}
\end{equation}

\subsection{Fierz identities}
\label{ss.fi}

In this article, we utilize the following Fierz identity and its
derivatives,
\begin{equation}
\theta_{MW} \Gamma^m \phi_{MW} \Gamma_m \psi_{MW} + \phi_{MW} \Gamma^m \psi_{MW} \Gamma_m \theta_{MW} + \psi_{MW} \Gamma^m \theta_{MW} \Gamma_m \phi_{MW} = 0 ,
\label{Fierz}
\end{equation}
where $\theta_{MW} , \phi_{MW} , \psi_{MW}$ are $10$ dim. Majorana-Weyl spinors.

For {\it double-decker} Weyl spinors,
\begin{eqnarray}
{V \!\!\!\! /}_{\Theta , \Phi} \Psi + {V \!\!\!\! /}_{\Phi , \Psi} \Theta + {V \!\!\!\! /}_{\Psi , \Theta} \Phi &=& 0 , \nonumber \\
\bar{\Psi} {V \!\!\!\! /}_{\Theta , \Phi} + \bar{\Theta} {V \!\!\!\! /}_{\Phi , \Psi} + \bar{\Phi} {V \!\!\!\! /}_{\Psi , \Theta} &=& 0 ,
\label{Fierzd}
\end{eqnarray}
where
\begin{eqnarray}
V_{\Theta , \Phi}^{m} &=& (\bar{\Theta} {\it 1} \Gamma^{m} \Phi) \ {\it 1} + (\bar{\Theta} I \Gamma^{m} \Phi) \ I \nonumber \\
 &=& (\bar{\theta} \Gamma^{m} \phi + \theta \Gamma^{m} \bar{\phi}) \ {\it 1} + (\bar{\theta} \Gamma^{m} \bar{\phi} + \theta \Gamma^{m} \phi) \ I \nonumber \\
 &=& \left[
\begin{array}{cc}
\bar{\theta} \Gamma^{m} \phi + \theta \Gamma^{m} \bar{\phi} & \bar{\theta} \Gamma^{m} \bar{\phi} + \theta \Gamma^{m} \phi \\
\bar{\theta} \Gamma^{m} \bar{\phi} + \theta \Gamma^{m} \phi & \bar{\theta} \Gamma^{m} \phi + \theta \Gamma^{m} \bar{\phi}
\end{array}
\right] .
\label{V}
\end{eqnarray}

Using (\ref{V}), one can show
\begin{equation}
d {V \!\!\!\! /} \Theta - 2 {V \!\!\!\! /} d \Theta = 0 \ , \ \bar{\Theta} d {V \!\!\!\! /} + 2 d \bar{\Theta} {V \!\!\!\! /} = 0 , 
\label{dVTheta}
\end{equation}
\begin{equation}
d {V \!\!\!\! /} d \Theta = 0 \ , \ d \bar{\Theta} d {V \!\!\!\! /} = 0 ,
\label{dVdTheta}
\end{equation}
\begin{equation}
d V \cdot V = V \cdot d V = 0 .
\label{dVV}
\end{equation}
Here we use the following abbreviation
\begin{equation}
V^{m} = V_{d \Theta , \Theta}^{m} .
\label{Vd}
\end{equation}

\section{Proof of Eq.(\ref{dKeIHepi})}
\label{s.pe}

In this subsection, we prove Eq.(\ref{dKeIHepi})
\begin{equation}
d \ [ K e^{- I \frac{H + \bar{H}}{2}} e^{\pm {\pi \!\!\! /}} ] = (\pm \frac{1}{4} d {V \!\!\!\! /}) K e^{- I \frac{H + \bar{H}}{2}} e^{\mp {\pi \!\!\! /}} + K e^{- I \frac{H + \bar{H}}{2}} e^{\pm {\pi \!\!\! /}} (\pm \frac{1}{4} d {V \!\!\!\! /}) . \nonumber
\end{equation}

Using the Clifford algebra $\{ \Gamma^{m} , \Gamma^{n} \} = 2 \eta^{mn}$,
we obtain 
\begin{equation}
\Gamma_{m} \ {\pi \!\!\! /}^{l} = (-)^{l} \ {\pi \!\!\! /}^{l} \Gamma_{m} + 2 l \ \pi_{m} \ {\pi \!\!\! /}^{l-1} ,
\label{Gammapil}
\end{equation}
and then
\begin{eqnarray}
 & & \Gamma_{m} e^{\pi \!\!\! /} \ \ - e^{- \pi \!\!\! /} \Gamma_{m} = \ \ 2 \pi_{m} e^{\pi \!\!\! /} \ \ = \ \ 2 e^{- \pi \!\!\! /} \pi_{m} , \nonumber \\
 & & \Gamma_{m} e^{- \pi \!\!\! /} - e^{\pi \!\!\! /} \ \ \Gamma_{m} = - 2 \pi_{m} e^{- \pi \!\!\! /} = -2 e^{\pi \!\!\! /} \ \ \pi_{m} .
\label{Gammaepi}
\end{eqnarray}

Also we obtain
\begin{eqnarray}
d {\pi \!\!\! /}^{l} &=& \ \ \ \ \ \ \ \ \ \ l \ {\pi \!\!\! /}^{l-1} \ d {\pi \!\!\! /} - \ \ \ \ \ \ \ \ \ \ l (l-1) \ {\pi \!\!\! /}^{l-2} \ \pi \cdot d \pi \nonumber \\
 &=& (-)^{l-1} \ l \ d {\pi \!\!\! /} \ {\pi \!\!\! /}^{l-1} + (-)^{l-2} \ l (l-1) \ d \pi \cdot \pi \ {\pi \!\!\! /}^{l-2} ,
\label{dpil}
\end{eqnarray}
and then
\begin{eqnarray}
 & & d e^{\pi \!\!\! /} \ \ = \ \ \frac{1}{2} d {\pi \!\!\! /} e^{- \pi \!\!\! /} + \frac{1}{2} e^{\pi \!\!\! /} \ \ d {\pi \!\!\! /} , \nonumber \\
 & & d e^{- \pi \!\!\! /} = - \frac{1}{2} d {\pi \!\!\! /} e^{\pi \!\!\! /} \ \ - \frac{1}{2} e^{- \pi \!\!\! /} d {\pi \!\!\! /} .
\label{depi}
\end{eqnarray}

Using Eqs. (\ref{Gammaepi}),(\ref{depi}) and 
\begin{equation}
d \pi^m = \frac{1}{2} Re ( d \theta \Gamma^{m} d {\bar \theta}) ,
\label{dpim}
\end{equation}
\begin{eqnarray}
d H^{(2)} &=& - d C^{2} = - R^{(3)} = - \pi \cdot d \theta \Gamma d \theta , \nonumber \\
d {\bar H}^{(2)} &=& - d {\bar C}^{(2)} = - {\bar R}^{(3)} = - \pi \cdot d {\bar \theta} \Gamma d {\bar \theta} ,
\end{eqnarray}
we can show Eq. (\ref{dKeIHepi}). 

\section{Useful Formulas}
\label{s.uf}

\subsection{definitions}
\label{ss.def}

\begin{equation}
\Theta_j = {V \!\!\!\! /}^j \Theta \ , \ {\bar \Theta}_j = {\bar \Theta} {V \!\!\!\! /}^j .
\label{Thetaj}
\end{equation}

\begin{eqnarray}
\Omega^{(\pm)} &=& \sum_{j=0} \frac{(\pm)^j}{2^j (2j+1)!!} \Theta_{j} , \nonumber \\
{\bar \Omega}^{(\pm)} &=& \sum_{j=0} \frac{(\pm)^j}{2^j (2j+1)!!} {\bar \Theta}_{j} .
\label{Omega}
\end{eqnarray}

\begin{eqnarray}
\Sigma^{(\pm)}_{\Delta} &=& \ \ \frac{1}{2} \sum_{j=0} \frac{j+1}{2^{j} (2j+3)!!} {V \!\!\!\! /}_{\Delta} (\pm)^j \Theta_{j} \nonumber \\
 & & \mp \frac{1}{4} \sum_{j=0} \frac{(j+2)(j+1)}{2^{j} (2j+5)!!} V_{\Delta} \cdot V  (\pm)^j \Theta_{j} , \nonumber \\
{\bar \Sigma}^{(\pm)}_{\Delta} &=& \ \ \frac{1}{2} \sum_{j=0} \frac{j+1}{2^{j} (2j+3)!!} (\pm)^j {\bar \Theta}_{j}  {V \!\!\!\! /}_{\Delta} \nonumber \\
 & & \mp \frac{1}{4} \sum_{j=0} \frac{(j+2)(j+1)}{2^{j} (2j+5)!!} (\pm)^j {\bar \Theta}_{j} V_{\Delta} \cdot V .
\label{Sigma}
\end{eqnarray}

\begin{eqnarray}
\Lambda^{(\pm)}_{\Delta_{1} , \Delta_{2}} &=& \ \ \frac{1}{8} \sum_{j=0} \frac{(j+2)(j+1)}{2^{j} (2j+5)!!} {V \!\!\!\! /}_{[ \Delta_{1}} {V \!\!\!\! /}_{\Delta_{2} ]} (\pm)^j \Theta_{j} \nonumber \\
 & & \mp \frac{1}{8} \sum_{j=0} \frac{(j+3)(j+2)(j+1)}{2^{j} (2j+7)!!} {V \!\!\!\! /}_{[ \Delta_{1}} V_{\Delta_{2} ]} \cdot V (\pm)^j \Theta_{j} \nonumber \\
 & & - \frac{1}{16} \sum_{j=0} \frac{(j+4)(j+3)(j+2)(j+1)}{2^{j} (2j+9)!!} V_{\Delta_{1}} \cdot V \ V_{\Delta_{2}} \cdot V (\pm)^j \Theta_{j} .
\label{Lambda}
\end{eqnarray}

In Eqs.(\ref{Sigma}),(\ref{Lambda}), we use the following abbreviation
\begin{equation}
V_{\Delta}^m = V_{\Delta \Theta \ , \ \Theta}^{m} .
\label{VDelta} 
\end{equation}

\subsection{formulas involving exterior differential $d$}
\label{ss.d}

\begin{eqnarray}
d \Theta_{j} &=& (-)^{j-1} \frac{2j+1}{2} d {V \!\!\!\! /} \Theta_{j-1} \ \ \ \ \ (j \geq 1) , \nonumber \\
d \Theta_{0} &=& d \Theta , \nonumber \\
d \bar{\Theta}_{j} &=& \frac{2j+1}{2} \bar{\Theta}_{j-1} d {V \!\!\!\! /} \ \ \ \ \ (j \geq 1) , \nonumber \\
d \bar{\Theta}_{0} &=& d \bar{\Theta} .
\label{dThetaj}
\end{eqnarray}

\begin{eqnarray}
d \Omega^{(\pm)} &=& d \Theta \pm \frac{1}{4} d {V \!\!\!\! /} \Omega^{(\mp)} , \nonumber \\
d {\bar \Omega}^{(\pm)} &=& d {\bar \Theta} \pm \frac{1}{4} {\bar \Omega}^{(\pm)} d {V \!\!\!\! /} .
\label{dOmega}
\end{eqnarray}

\subsection{formulas involving variation $\Delta$}
\label{ss.delta}

Under even variation $\Delta \theta$,$\Delta {\bar \theta}$
\begin{eqnarray}
\Delta \Theta_{j} &=& \frac{2j+1}{2} \Delta {V \!\!\!\! /} \ \Theta_{j-1} - (j+1)(j-1) \Delta V \cdot V \Theta_{j-2} - {V \!\!\!\! /}^{j-1} \varrho \ \ \ \ \ (j \geq 2) , \nonumber \\
\Delta \Theta_{1} &=& \frac{3}{2} \Delta {V \!\!\!\! /} \ \Theta - \varrho , \nonumber \\
\Delta \Theta_{0} &=& \Delta \Theta , \nonumber \\
\Delta {\bar \Theta}_{j} &=& \frac{2j+1}{2} {\bar \Theta}_{j-1} \Delta {V \!\!\!\! /} - (j+1)(j-1) {\bar \Theta}_{j-2} V \cdot \Delta V -  {\bar \varrho} {V \!\!\!\! /}^{j-1} \ \ \ \ \ (j \geq 2) , \nonumber \\
\Delta \Theta_{1} &=& \frac{3}{2} {\bar \Theta} \Delta {V \!\!\!\! /} - {\bar \varrho} , \nonumber \\
\Delta {\bar \Theta}_{0} &=& \Delta {\bar \Theta} ,
\label{deltaThetaj}
\end{eqnarray}
where
\begin{eqnarray}
\varrho &=& \frac{1}{2} d {V \!\!\!\! /}_{\Delta} \Theta - {V \!\!\!\! /}_{\Delta} d \Theta , \nonumber \\
{\bar \varrho} &=& \frac{1}{2} {\bar \Theta} d {V \!\!\!\! /}_{\Delta} - d {\bar \Theta} {V \!\!\!\! /}_{\Delta} .
\label{varrho}
\end{eqnarray}

Using
\begin{eqnarray}
\Delta {V \!\!\!\! /} = d {V \!\!\!\! /}_{\Delta} + 2 {V \!\!\!\! /}_{d \Theta , \Delta \Theta} ,
\label{deltaV}
\end{eqnarray}
\begin{eqnarray}
\Delta V \cdot V &=& - d [ V_{\Delta} \cdot \Theta ] + 2 V_{\Delta} \cdot d V ,
\nonumber \\
V \cdot \Delta V &=& d [ V \cdot V_{\Delta} ] -2 d V \cdot V_{\Delta} ,
\label{deltaVV}
\end{eqnarray}
\begin{eqnarray}
\varrho &=& \frac{1}{2} d [ {V \!\!\!\! /}_{\Delta} \Theta ] - \frac{3}{2} {V \!\!\!\! /}_{\Delta} d \Theta , \nonumber \\
{\bar \varrho} &=& \frac{1}{2} d [ {\bar \Theta} {V \!\!\!\! /}_{\Delta} ] - \frac{3}{2} d {\bar \Theta} {V \!\!\!\! /}_{\Delta} ,
\label{varrho2}
\end{eqnarray}
we obtain
\begin{eqnarray}
\Delta \Omega^{(\pm)} &=& \Delta \Theta \pm \frac{1}{2} {V \!\!\!\! /}_{d \Theta , \Delta \Theta} \Omega^{\pm} - \frac{1}{4} d {V \!\!\!\! /} \ \Sigma^{(\pm)}_{\Delta} \pm d \Sigma^{(\mp)}_{\Delta} , \nonumber \\
\Delta {\bar \Omega}^{(\pm)} &=& \Delta {\bar \Theta} \pm \frac{1}{2} {\bar \Omega}^{(\pm)} {V \!\!\!\! /}_{d \Theta , \Delta \Theta} + \frac{1}{4} {\bar \Sigma}^{(\pm)}_{\Delta} d {V \!\!\!\! /} \pm d {\bar \Sigma}^{(\pm)}_{\Delta} ,
\label{deltaOmega}
\end{eqnarray}
and
\begin{equation}
\Delta_{[2} \Sigma^{(\pm)}_{\Delta_{1}]} = \frac{1}{2} {V \! \! \! \! /}_{\Delta_{1}, \Delta_{2}} \ \Omega^{(\pm)} \pm \frac{1}{2} {V \! \! \! \! /}_{[\Delta_{2}, d \Theta} \ \Sigma^{(\pm)}_{\Delta_{1}]} + \frac{1}{4} d {V \! \! \! \! /} \ \Lambda^{(\pm)}_{\Delta_{1}, \Delta_{2}} \pm d \Lambda^{(\mp)}_{\Delta_{1}, \Delta_{2}} .
\label{deltaSigma}
\end{equation}

\section{Other Conventions}
\label{s.oc}

\subsection{exterior differential}
\label{ss.ed}

Exterior differential is defined so as to obey the following Leibniz
rule
\begin{equation}
d (\omega^{(p)} \omega^{(q)}) =  \omega^{(p)} d (\omega^{(q)}) + (-)^q d (\omega^{(p)}) \omega^{(q)} ,
\label{Leibniz}
\end{equation}
where $\omega^{(p)}$ denotes a $p$ form.

\subsection{complex conjugation}
\label{ss.cc}

Following \cite{CW}, we adopt the convention that the complex
conjugation of the product of fermionic quantities is defined without a
reverse of the order.

\subsection{(anti-) symmetrizing symbol}
\label{ss.ss}

The symbol $\{ ... \}$ or $[ ... ]$ symmetrizes or anti-symmetrizes $n$ indices
inside it without the normalization factor $(n!)^{-1}$.

%

\end{document}